\documentstyle[11pt,newpasp,epsf,a4wide]{article}
\newcommand{\ds}{\displaystyle}
\newcommand{\inv} {\frac {1}}

\newcommand{\eqn} [1] {
\begin{equation}#1
\end{equation}}
\newcommand{\eqna} [1] {
\begin{eqnarray}#1
\end{eqnarray}}

\newcommand{\fig}[3]{
      \begin{figure}[tbp]
\epsfxsize=13cm
\begin{center}
\epsfbox{#1}
\end{center}
	\caption{#2}
	\label{#3}
        \end{figure} }
\begin{document}

\title{Characterizing the dynamic properties of the solar turbulence with 3-D simulations: \\Consequences in terms of $p$-mode excitation.}

\author{Samadi R.\altaffilmark{1}}
\affil{Astronomy Unit, Queen Mary, University of London, London, UK.\\
LESIA, Observatoire de Paris, , Meudon, France.}

\author{Nordlund A.}
\affil{Niels Bohr Institute for Astronomy, Physics and Geophysics, Copenhagen, Denmark.}

\author{Stein R.F.}
\affil{Michigan State University, USA}

\author{Goupil M.-J.}
\affil{LESIA, Observatoire de Paris, Meudon, France.} 

\altaffiltext{1}{Postal address : LESIA Observatoire de Paris-Meudon, 
Bat. 12, 5 place J. Janssen, F-92195 Meudon, France. \\e-mail: Reza.Samadi@obspm.fr }

\author{Roxburgh I.}
\affil{Astronomy Unit, Queen Mary, University of London, London, UK.\\
LESIA, Observatoire de Paris, Meudon, France.}

\begin{abstract}
A 3D simulation of the upper part of the solar convective zone is used to derive constraints about the averaged and dynamic properties of solar turbulent convection.  
Theses  constraints are then used to compute the acoustic energy supply rate $P$ injected into the solar radial oscillations  according to the theoretical  expression in  Samadi \& Goupil (2001). The result is compared with solar seismic data.
\\Assuming, as it is usually  done,    a gaussian model for the frequency component $\chi_k(\nu)$ of the model of turbulence, it is  found that the computed  $P(\nu)$ is underestimated compared with the solar seismic data by a factor $\sim 2.5$. 
\\A frequency analysis of the solar simulation shows that the gaussian model indeed does not correctly model  $\chi_k(\nu)$ in the frequency range where the acoustic energy injected into the solar $p$-modes is important  ($\nu \simeq 2 - 4$~mHz). One must consider an additional non-gaussian component for $\chi_k(\nu)$ to reproduce its behavior.
Computed values  of $P$ obtained with this non-gaussian component reproduce better the solar seismic observations. 
This non-gaussian component leads to a Reynolds stress contribution of the same order than the one arising from  the advection of the turbulent fluctuations of entropy by the turbulent motions. 
\end{abstract}

\section{Introduction}

Turbulent motions in the upper convective zone of solar like stars,  generate acoustic energy, which in turn is injected at the rate $P(\nu)$  into $p$-modes oscillations ($\nu$ is the frequency of a given mode). Solar-type oscillations are therefore meant as stochastically excited by turbulent convection. This type of excitation process concerns low massive stars with an outer convective zone.

Measurement of the  rate of acoustic energy, $P(\nu)$, injected into solar-like oscillations is one of the goals of future space seismic missions COROT (Baglin et al 1998) and Eddington (Favata et al 2000).
These seismic data will  then make it possible to constrain the theory of 
excitation and damping of solar-type oscillations, which in turn will provide valuable information
about the properties of stellar convection zones.

Theoretical formulations for $P(\nu)$ offer the advantage of testing {\it separately} several properties entering in excitation mechanism.
Here we consider the formulation by Samadi \& Goupil (2001, see section~\ref{formulation} and also Samadi (2001)  for a detailed summary).  
%)  is proposed in Samadi \& Goupil (2001)  which generalizes and supplements previous theories ( One advantage of this formulation is the possibility of studying consistently the effect of using different turbulent spectra on the oscillation amplitudes and their frequency-dependence (see section~\ref{formulation}). 

Computation of  $P(\nu)$  requires an accurate knowledge of the dynamic properties of turbulence in stars. Unfortunately, the current observations of the solar granulation  cannot provide a precise enough description of the turbulent spectrum properties.

In the present work we use a 3D simulation of the upper part of the solar convective zone to determine the dynamic properties of solar turbulence (see section~\ref{the3Dconstraints}) which are necessary to compute $P(\nu)$.
Our calculations are then compared with  the helioseismic  constraints from the GOLF/SOHO instrument (see section~\ref{consequences}).

\section{A  theoretical formulation for $P(\nu)$}
\label{formulation}

The calculation of the rate $P(\nu)$  at which a given $p$-mode is excited results from an integration over the stellar mass ($m$) and local integrations over  distance ($r$) and time ($t$) of the mode eigenfunction ($\xi$) and the correlation product of the excitation source ($<\vec S \, \vec S>$). This expression can be written in a schematic form as :
\eqn{
P(\nu) \propto \int d{\rm m} \, \int d{\rm r} \, d\tau \, \, \vec \xi \, .  \,< \vec S \, \vec S > \, .  \,\vec \xi
}

The excitation source ($\vec S$) has two identified origines : the turbulent Reynolds stress and the advection of the turbulent entropy fluctuations by the turbulent motions.
$<\vec S \, \vec S>$  is expressed in terms of the turbulent kinetic energy spectrum $E(k,\nu)$ and the spectrum of the entropy fluctuations  $E_s(k, \nu)$ where $k$ is the wavenumber of a given turbulent element.
Detailled expressions for $P(\nu)$ and $< \vec S \, \vec S >$  are given in Samadi \& Goupil (2001).

Following Stein (1967), $E(k,\nu)$ is split into a spatial component $E(k)$ and a frequency component $\chi_k(\nu)$ as
\eqna{
E( k,\nu) =E(  k) \, \chi_k(\nu) \; .
\label{eqn:ek_chik}
}
The same decomposition is assumed for $E_s(k, \nu)$.

A gaussian shape for $\chi_k(\nu)$ is usually assumed in the calculation of  $P(\nu)$ (e.g. Goldreich \& Keeley 1977). This is equivalent to suppose that two distant points in the turbulent medium are uncorrelated.
In section~\ref{the3Dconstraints} below, the gaussian hypothesis is compared with the properties of $\chi_k(\nu)$ inferred from a 3D simulation of the solar convective zone. Section~4 discusses consequences of either choice on the excitation rate $P(\nu)$.

\section{Constraints from a 3D simulation of the upper part of the solar convective zone.}
\label{the3Dconstraints}

We study a 3D simulation of the upper part of the solar convective zone obtained with the 3D numerical code developed at the Niels Bohr Institute for Astronomy, Physics and Geophysics (Copenhagen, Denmark). Physical assumptions are described in Stein \& Nordlund (1998).

The simulated domain is 3.2 Mm deep and its surface is 6 x 6 ${\rm Mm}^2$. The grid of mesh points is 256 x 256 x 163, the total duration 27 mn and the sampling time 30s.

The simulation data are used to determine the quantities $E(k,z)$, $E_s(k,z)$ and $\chi_k(\nu,z)$ involved in the theoretical expression for the excitation rate $P(\nu)$.

We proceed in two steps (more details will be given in Samadi et al  2002a and Samadi et al 2002b):
\begin{itemize}
\item We  compute at each layer $z$ the 2D Fourier transform, along horizontal plans, of the velocity field $\vec u$ and the entropy $s$ and perform integrations over circles with radius $k$. This provides ${\hat {\bf u}}(k,z,t)$ and $\hat s(k,z,t)$ where  ${\bf k}$ is the wavenumber along the horizontal plan. We finally time average each quantities ${\hat {\bf u}}$ and $\hat s$ over the time series. This provides $ { \hat {\bf u}}(k,z)$,  ${\hat s}(k,z)$ and then  time averaged kinetic energy spectrum   $E(k) \equiv  {\hat {\bf u}} ^2 (k,z)$ and the time averaged spectrum of the entropy fluctuations $E_s(k) \equiv \hat s ^2 (k,z)$.
\item At 3 different layers of the simulated domain, we compute the 3D Fourier transform, with respect to time and along the horizontal plan,  of the velocity field ${\bf u}$. We next  perform integrations over circles with radius $k$. This yields $ {\hat {\bf u}}(k,z,\nu)$ and therefore $E(k,\nu,z) \equiv  {\hat {\bf u}}^2(k,\nu,z)$  and - using Eq.(2) -  $\chi_k(\nu,z)$.
\end{itemize}

The dependence of $\chi_k$ with frequency is plotted at the top of the superadiabatic region -where the $p$-modes excitation is the largest - (Fig.~1), 0.3 Mm deeper (Fig.~2) and 0.6 Mm deeper (Fig.~3).

\fig{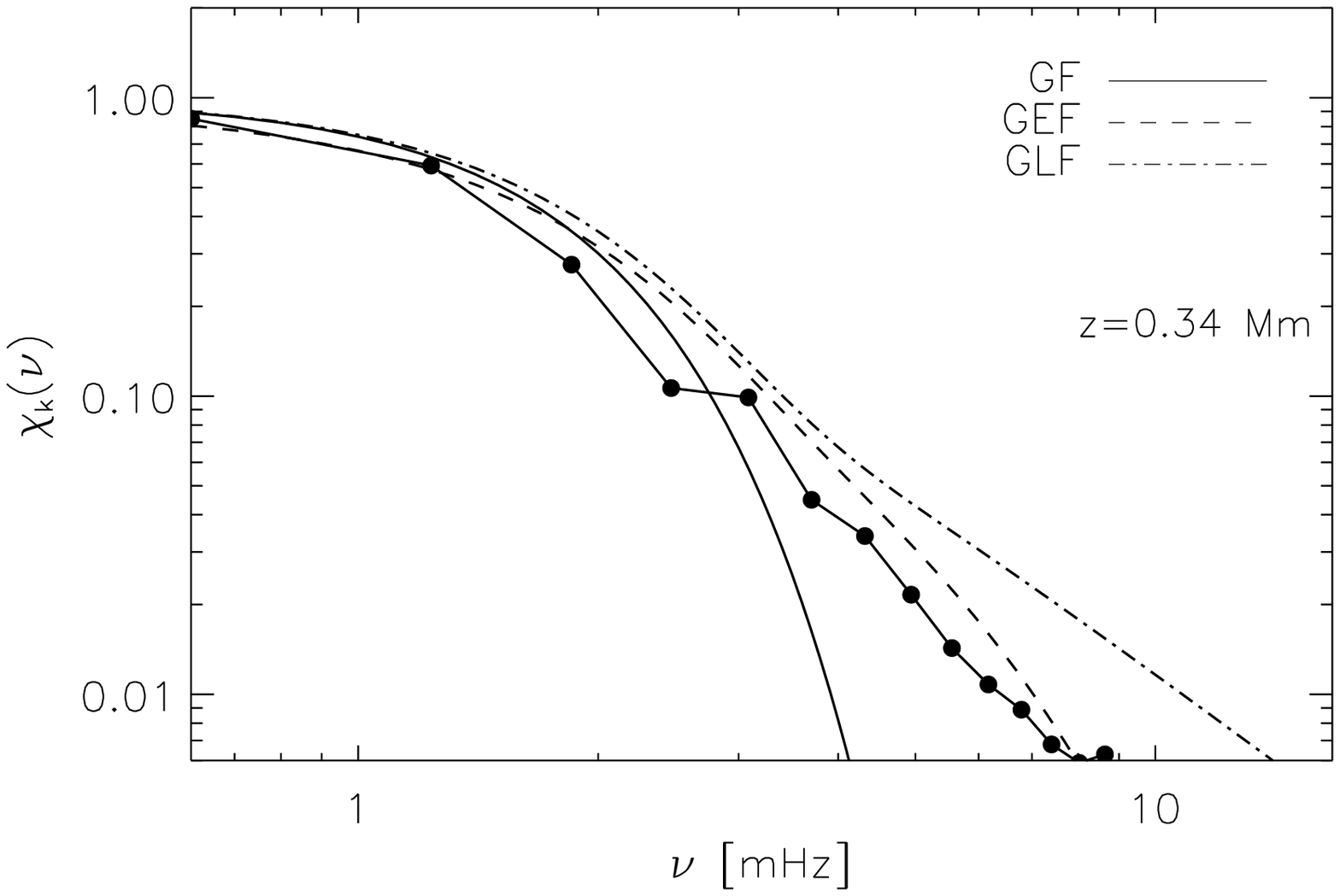}{The solid curve with dots represents  $\chi_k(\nu)$ obtained from the simulation at the top of the superadiabatic region ($z=0.04$~Mm) and for the wavenumber $k$ at which $E(k)$ is maximum. The solid curve represents the Gaussian function (GF, Eq.~\ref{eqn:GF}), the dashed curve the Gaussian Exponential function (GEF, Eq.~\ref{eqn:GEF}) and the dots-dashed curve the  Gaussian Lorentzian function (GLF, Eq.~\ref{eqn:GF}).}{}
\fig{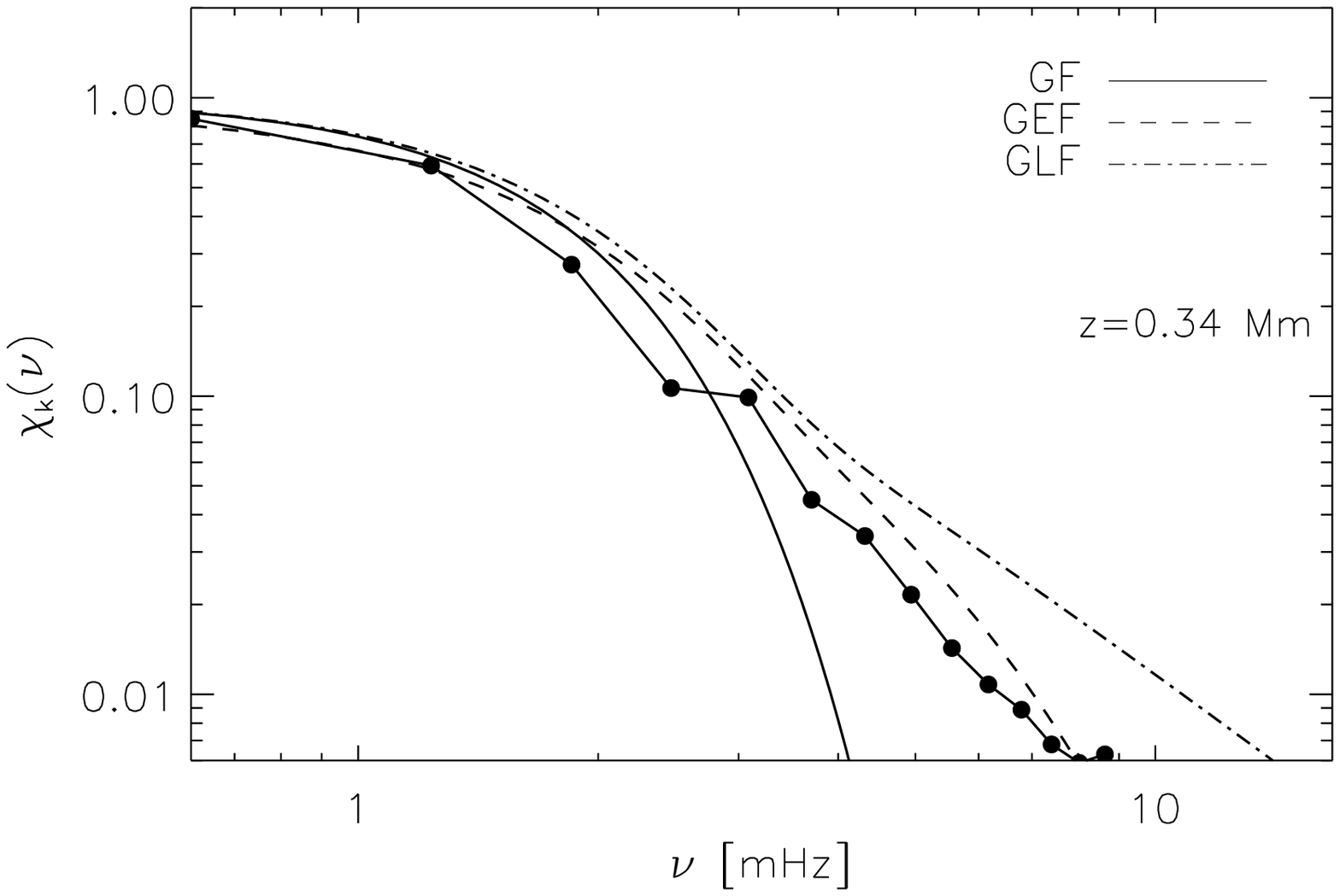}{Same as Fig 1, at a deeper layer (z=0.34 Mm).}{}
\fig{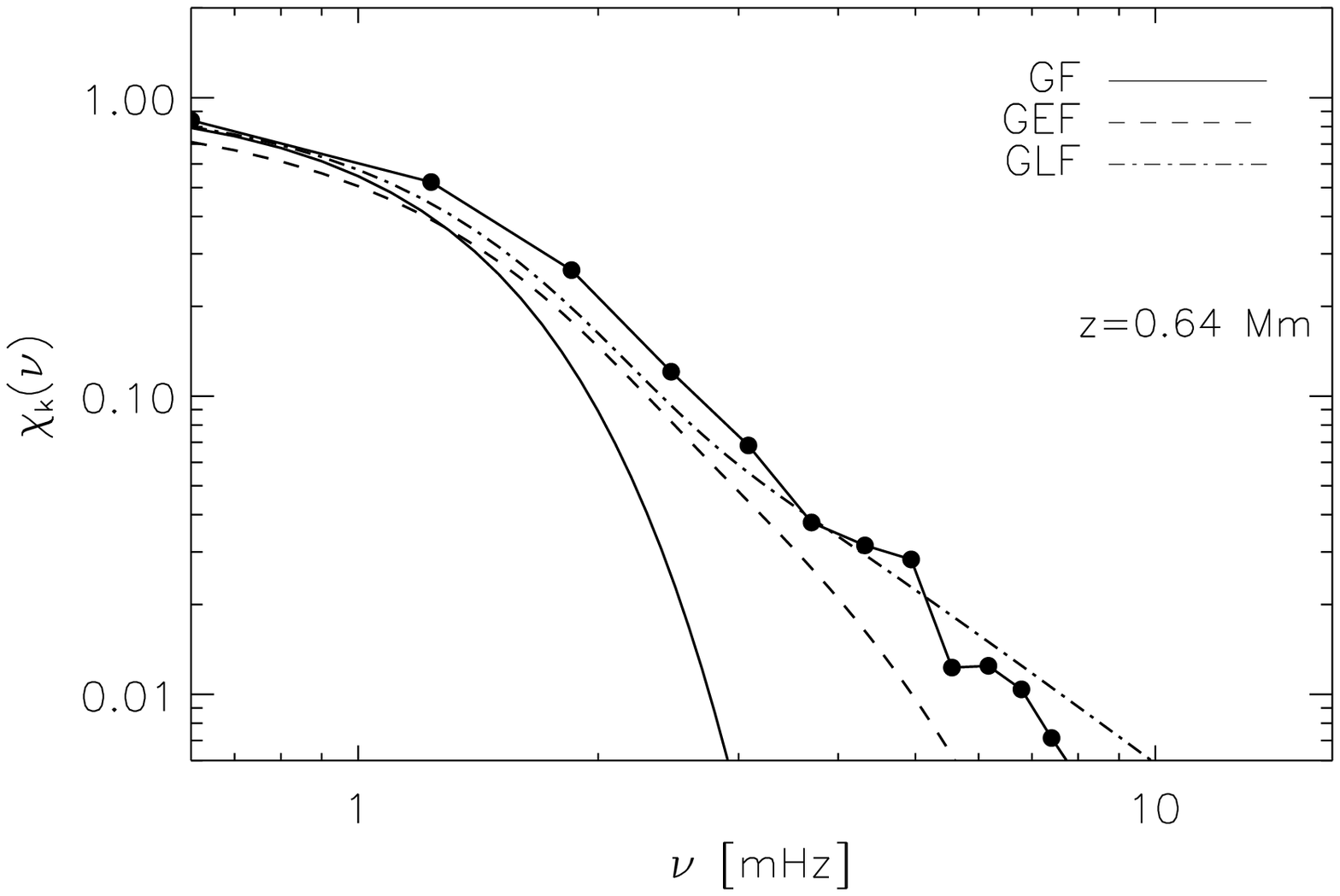}{Same as Fig 1, at a deeper layer (z=0.64 Mm).}{}

\vspace{0.8cm}

We compare $\chi_k(\nu)$ obtained from the simulation analysis with three analytical functions. These are the Gaussian function (GF hereafter):
\eqn{
\chi_k (\nu ) = \inv  { \nu_k \, \sqrt{\pi}}  e^{-(\nu / \nu_k)^2} \; ,
\label{eqn:GF}
}  the Gaussian  plus an Exponential function  (GEF hereafter): 
\eqn{
\chi_k (\nu ) = {1 \over 2 } \, \left (  \inv  { \nu_k \, \sqrt{\pi}}  e^{-(\nu / \nu_k)^2} + \frac{1}{2 \nu_k}  e ^{-| \nu/\nu_k | } \right ) \;  ,
\label{eqn:GEF}
} and the Gaussian  plus a Lorentzian function (GLF hereafter):
\eqn{
\chi_k (\nu ) = {1 \over 2 } \, \left (  \inv  { \nu_k \, \sqrt{\pi}}  e^{-(\nu / \nu_k)^2} + \frac{1}{\pi \, \nu_k}\, \frac{1}{1+\left (\nu/\nu_k\right)^2} \right ) 
\label{eqn:GLF}
}
where $\nu_k$ is the line-width at half maximum of $\chi_k$. 
$\nu_k$ is related to $\tau_k$  the characteristic correlation time-scale of an eddy with wavenumber $k$  as 
\eqn{
\nu_k \equiv {  (\pi \tau_k) ^{-1}}
}
and $\tau_k$ is related to the velocity $u_k$ of  an eddy with wavenumber $k$  as
\eqn{
\tau_k \equiv \lambda \, (k \, u_k)^{-1} \; .
\label{eqn:tauk}
}
The velocity $u_k$ is obtained from the kinetic energy spectrum $E(k)$ (Stein 1967) as
\eqn{
u_k^2 =  \int_k^{2 k}\,{\rm d}k\,E(k)\,.
\label{eqn:uk2}
}
The parameter $\lambda$ in Eq.~\ref{eqn:tauk} accounts for our lack of precise knowledge of the  time correlation  $\tau_k$ (or $\nu_k$) in stellar conditions.
We find however tgat in most part of the excitation region, the line width of $\chi_k$ is   satisfactorily reproduced with $\lambda=1$.

At the top of the superadiabatic region, the GF and GLF do not correctly model $\chi_k(\nu)$ whereas the GEF is in better agreement (see Fig.~1). 
However the discrepancies between the GF (or the GLF) and the simulation data occur mostly above the solar cut-off frequency ($\nu \sim 5.5$~mHz). Discrepancies between the GF (or the GLF) and the simulation data have then minor consequences for the $p$-modes excitation in this region.

This is not the case deeper in the simulation where the largest discrepancies  between the GF and data occur in the frequency range where the larger amount of acoustic energy is injected into the $p$-modes ($\nu \sim 2 - 4$~mHz).
The GEF  and  GLF  reproduce better  than the GF the $\nu$-variation of  $\chi_k$ below the top of the superadiabatic region ($z < 0$~Mm, see Fig.~2~\&~3).

\section{Consequences in terms of $p$-mode excitation}
\label{consequences}

We compute $P(\nu)$ according to Eq.~(1): 
\begin{itemize}
\item  The eigenfunctions ($\xi$) and their frequencies ($\nu$) are computed with Balmforth's (1992) non-adiabatic code for a solar 1D mixing-length model based on Gough's (1977) non-local time-dependent formulation of convection.
\item The $k$-dependency of $E(k,z)$,  is modeled as following :
\eqn{
\begin{array}{cccc}
 E(k) \propto (k/k_0)^{+1}   &  \textrm{for} & k_0  >  k > 0.17 \, k_0\\
 E(k) \propto  (k/k_0)^{-5/3}  &   \textrm{for} &  k > k_0                \\
\end{array}
}
where $k_0 = 2 \pi /  \beta \Lambda$, $\Lambda=\alpha H_p$ is the mixing-length, $H_p$ the pressure scale height and $\alpha$ the mixing-length parameter. The value of $\alpha$ is this of the 1D solar model for consistency.The value of $k_0$ hence of $\beta$ is obtained from the simulation.  This analytical $k$-dependency of $E$ reproduces the global features of $E$ arising from the simulation. Same model is  considered for  $E_s(k,z)$.
\item $E(k,z)$ and $E_s(k,z)$ verifie the normalization conditions: \eqn{
\begin{array}{lll}
\vspace{0.2cm}
\ds \int  {\rm d} k  \, E(k,z) & =  & \ds  {1 \over 2} \, < {\bf u}^2- <{\bf u}>^2>(z)   \\
\vspace{0.2cm}
\ds \int  {\rm d} k  \, E_s(k,z)   & =  & \ds {1 \over 2} \, <s^2-  <s>^2>(z) 
\end{array}
\label{eqn:EEzEs}
}
where $< . >$ denotes time and horizontal  average.
The total energy contained in  $E(k,z)$ and $E_s(k,z)$ and their depth dependences are then obtained from the simulation according to Eq.(\ref{eqn:EEzEs}). 
\item For the frequency component $\chi_k(\nu)$,  we assume successively the GF, the GEF and the GLF (see section~\ref{the3Dconstraints}).
\end{itemize}

\fig{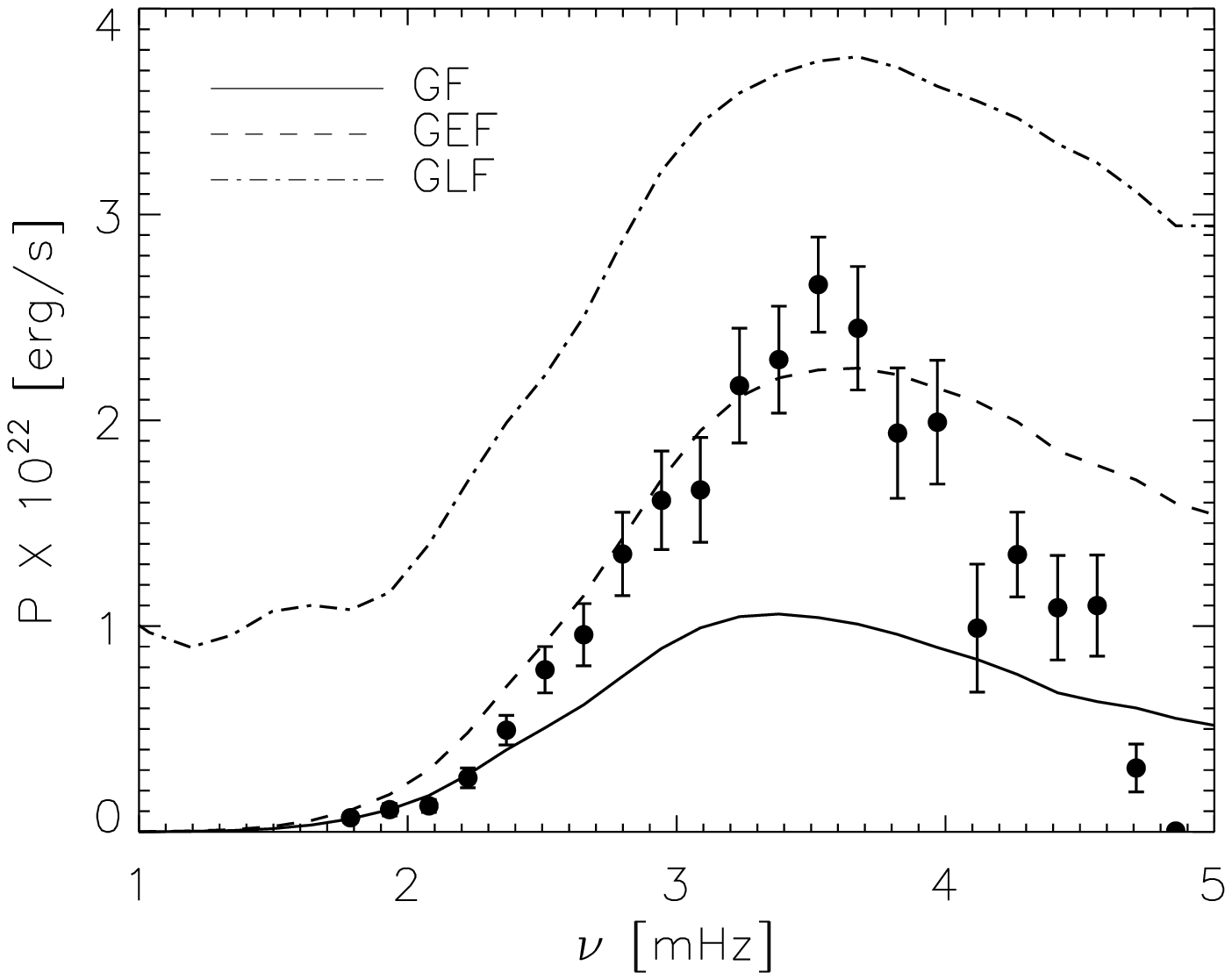}{The curves correspond to computed  $P(\nu)$ in which we assume different analytical functions for $\chi_k(\nu)$ : the GF (solid curve), the GEF (dashed curve) and the GLF (dots-dashed curve). The dots represent  $P(\nu)$ derived from the amplitudes and line widths of the $\ell=0$ $p$-modes measured by GOLF/SOHO instrument and kindly provided by F. Baudin (Baudin et al 2003, see also Thiery et al 2000). No significant difference for our purpose here are observed between the GOLF/SOHO data and Libbrecht (1988) observations. }{}
 
Results are presented in Fig 4.
Using the GF leads to a significant over-estimation of $P(\nu)$. This is a consequence of the large discrepancy between the GF and the 3D simulation frequency dependence of $\chi_k(\nu)$.
Using the GLF yields a power $P(\nu)$ which is much larger than the observations in particular at high frequency. This is because at the top of the superadiabatic region - where the excitation is the largest - the GLF  over-estimates $\chi_k(\nu)$.

The best agreement between computed and observed  $P(\nu)$ is found when assuming the GEF. 
This is a direct consequence of the rather good agreement between the GEF and the dynamic behavior of the solar turbulence inferred from the simulation (see Sect.~3).

\section{Conclusions}

 Our investigation demonstrates the non-gaussian character of the stochastic excitation of solar $p$-modes. 
Indeed,  the gaussian function (GF) as a model for $\chi_k(\nu)$ leads to an important over-estimation of $P(\nu)$ whereas  the Gaussian Exponential Function (GEF), which decreases more slowly with  $\nu$ than the GF does,  results in a better agreement between  $P(\nu)$ and the seismic solar observations.
The maximum value of $P(\nu)$ is now  reproduced without any adjustement of free parameters  in contrast with previous approaches.

 The non-gaussian character of stochastic excitation causes the Reynolds stress contribution to be of the same order as the contribution arising from  the advection of the turbulent fluctuations of entropy by the turbulent motions (not shown here, see Samadi et al 2002b). 
This last result is  in better agreement with recent results by Stein \& Norlund (2001)  and contrasts with  previous results by Samadi et al (2001) based on the gaussian assumption for $\chi_k(\nu)$ and  also with  the estimation carried out by Goldreich et al (1994).

% In turn our work shows   the rather good consistency between the properties of the turbulent solar medium inferred from this 3D simulation and  the helioseismic observations.

 \acknowledgments

RS acknowledges support by the Particle Physics and Astronomy Research Council of the UK under the grant PPA/G/O/1998/00576.
We thank Guenter Houdek for providing us the solar model and its eigenfunctions and Frederic Baudin for providing us helioseismic measurements from the GOLF/SOHO instrument.

%--------------------------------------------------------------------
% BIBLIOGRAPHY
%--------------------------------------------------------------------


\begin{references}
\reference Baglin A. and The COROT Team, 1998, in IAU Symp. 185: New Eyes to See Inside the Sun and Stars, vol. 185, 301.
\reference Balmforth N.J., 1992, MNRS, 255, 603.
\reference Baudin et al, 2003, in preparation.
\reference Favata F., Roxburgh I., Christensen-Dalsgaard J., 2000, in The Third
  MONS Workshop : Science Preparation and Target Selection, 49-54.
\reference Goldreich P., Keeley D., 1977, ApJ, 212, 243.
\reference Goldreich P., Murray N., Kumar P., 1994, ApJ, 424, 466.
\reference Gough D., 1977, ApJ, 214, 196.
\reference Libbrecht K.G., 1988, ApJ, 334, 510.
\reference Samadi R. \& Goupil M.J., 2001, A\&A, 370, 136.
\reference Samadi R., Goupil MJ, Lebreton Y., 2001a, A\&A, 370, 147.
\reference Samadi R., 2001, in SF2A-2001: Semaine de l'Astrophysique Francaise, E148.
\reference Samadi R., Nordlund A., Stein R.F. , Goupil M.J., Roxburgh I., 2002a,''Numerical constraints on the model of stochastic excitation  of solar-type oscillations'' , in preparation, to be submitted to A\&A.
\reference Samadi R., Nordlund A., Stein R.F. , Goupil M.J., Roxburgh I., 2002b, ``Non gaussian character of stochastic excitation of solar $p$-modes'', in preparation, to be submitted to A\&A.
\reference Stein R.F., 1967, Solar Physics, 2, 385.
\reference Stein R.F., Nordlund A., 1998, ApJ, 499, 914.
\reference Stein R.F.; Nordlund A., 2001, ApJ, 546 , 583.
\reference Thiery S. et al, 2000, A\&A, 355, 743.
\end{references}
\end{document}